\documentclass[aps,prl,twocolumn,groupedaddress,nofootinbib,showpacs]{revtex4}
\usepackage[dvips]{graphicx}
\newif\ifdraft
\drafttrue
\newif\ifpreprint
\preprinttrue
\def\eqn#1{eq.~({\ref{#1}})}

\def\spa#1.#2{\left\langle#1\,#2\right\rangle}
\def\spb#1.#2{\left[#1\,#2\right]}

\def\tab#1{table~{\ref{#1}}}

\def\eps{\epsilon}

\newcommand{\eq}{\begin{equation}}
\newcommand{\eqe}{\end{equation}}
\newcommand{\eqa}{\begin{eqnarray}}
\newcommand{\eqae}{\end{eqnarray}}

\newcommand{\p}{\partial}
\newcommand{\teta}{\tilde{\eta}}

\newbox\charbox
\newbox\slabox
\def\s#1{{      % Feynman slash
        \setbox\charbox=\hbox{$#1$}
        \setbox\slabox=\hbox{$/$}
        \dimen\charbox=\ht\slabox
        \advance\dimen\charbox by -\dp\slabox
        \advance\dimen\charbox by -\ht\charbox
        \advance\dimen\charbox by \dp\charbox
        \divide\dimen\charbox by 2
        \raise-\dimen\charbox\hbox to \wd\charbox{\hss/\hss}
        \llap{$#1$}
}}

\begin{document}

\title{
\ifpreprint
\hbox{\rm \small
CERN-PH-TH/2012-254$\null\hskip 4.3cm\null$ \hfill 
MCTP-12-22$\null\hskip 4.3cm\null$ \hfill 
Saclay--IPhT--T12/076 \break}
\fi
Equivalent $D=3$ Supergravity Amplitudes from Double Copies \\ of Three-Algebra and Two-Algebra Gauge Theories}
 
\author{Yu-tin~Huang${}^{a,b}$ and Henrik~Johansson${}^{c,d}$
}

\affiliation{
${}^a$Michigan Center for Theoretical Physics, Department of Physics, University of Michigan, Ann Arbor, MI 48109, USA \\
${}^b$Department of Physics and Astronomy, UCLA, Los Angeles, CA
90095-1547, USA \\
${}^c$Theory Division, Physics Department, CERN, CH-1211 Geneva 23, Switzerland\\
${}^d$Institut de Physique Th\'eorique, CEA--Saclay,  F--91191
Gif-sur-Yvette cedex, France}

\begin{abstract} 
We show that three-dimensional supergravity amplitudes can be obtained as double copies of either three-algebra super-Chern-Simons matter theory or that of two-algebra super-Yang-Mills theory, when either theory is organized to display the color-kinematics duality. We prove that only helicity-conserving four-dimensional gravity amplitudes have nonvanishing descendants when reduced to three dimensions; implying the vanishing of odd-multiplicity S-matrix elements, in agreement with Chern-Simons matter theory. We explicitly verify the double-copy correspondence at four and six points for $\mathcal{N}=12,10,8$ supergravity theories and discuss its validity for all multiplicity. 
\end{abstract}

\pacs{04.65.+e, 11.15.Bt, 11.30.Pb, 11.55.Bq \hspace{1cm}}

\maketitle
It has been a long held view that four-dimensional perturbative quantum gravity defined by the Einstein-Hilbert action, with its supersymmetric completion, is ill defined due to the expectation of a proliferation of new divergences at each order of perturbation theory. In recent years such view has been challenged due to explicit calculations in $D\ge4$ dimensions~\cite{GravityThree, GravityFour, GravityThree2} showing the absence of previously expected divergences in $\mathcal{N}=8$ supergravity~\cite{N8Sugra}, as well as $\mathcal{N}=4$ supergravity~\cite{N4Sugra} where candidate counterterms that satisfy all known symmetries of the theory has been explicitly constructed~\cite{VanishingVolume}. 

The absence of divergences should imply the existence of either a hidden symmetry which the would-be-counterterm violates~\cite{Ferrara}, or hidden structures of perturbative amplitudes that results in tamer ultraviolet (UV) behavior. A proposal for the latter was given by Bern, Carrasco and one of the current authors (BCJ)~\cite{BCJ,BCJ2}, who conjectured that super-Yang-Mills (sYM) amplitudes can be reorganized such that the kinematic structure mirrors the Lie-algebra relations satisfied by the color group. Furthermore, it was proven~\cite{Square} that once such a {\it color-kinematics-dual} representation is found, gravity amplitudes can be simply obtained by taking a  {\it double copy} (squaring) of the duality-satisfying kinematic factors~\cite{BCJ}.  While this gauge-gravity relation is classically equivalent to the field-theory version of the Kawai-Levellen-Tye (KLT) open-closed string theory relations~\cite{KLT}, its seamless extension to loop level is unrivaled. Renewed analysis has shown that the double-copy relationship between gauge and gravity theories is intimately tied to improved UV behavior of maximal~\cite{BCJ2, BCJ3} and half-maximal supergravity theories~\cite{New}. 

The notion of a duality between color and kinematics is surprisingly universal -- it might well be a fundamental principle of nature. Besides being present in a wide range of $D$-dimensional Yang-Mills theories~\cite{BCJ}, the same structure has been observed in string theory~\cite{monodromy,StringTheoryStructures}, and in three-dimensional  Chern-Simons matter (CSm) theory~\cite{McLoughlin}. The latter case is the topic of this Letter.

In three dimensions, there are two important classes of gauge theories:  (super-)YM and (superconformal) CSm theory. Unlike the case of sYM theory, where the color dependence is governed by structure constants of a Lie two-algebra, the color structure for CSm is governed by a Lie three-algebra~\cite{N=8,N=6,N=5}.
While CSm theory can be equivalently formulated using ordinary Lie two-algebra, the properties relevant to this paper are better understood in the three-algebra formulation.
Recently Bargheer, He and McLoughlin~\cite{McLoughlin} showed that the amplitudes for the $\mathcal{N}=8$ CSm, also known as  Bagger-Lambert-Gustavsson (BLG) theory~\cite{N=8}, can be rearranged such that the kinematic parts mirror the relations of the three-algebra color structure. They demonstrated, at four and six points, by squaring the kinematic factors one obtains the amplitudes of the $\mathcal{N}=16$ supergravity theory constructed by Marcus and Schwarz~\cite{E8}. 

In this Letter, we show that tree-level amplitudes of CSm and sYM theories, via respective three- and two-algebra double copy relations, gives identical supergravity tree amplitudes -- clarifying the results of ref.~\cite{McLoughlin}. This result is consistent with the statement~\cite{E8} that the $\mathcal{N}=16$ Marcus-Schwarz theory is equivalent to the dimensional reduction of four-dimensional $\mathcal{N}=8$ supergravity, at least for the on-shell S-matrix. The equivalence of the two double copies is striking since, in contrast to CSm, odd-multiplicity S-matrix elements of sYM are nonvanishing. We show that R-symmetry constrains imply that the KLT relations give vanishing odd-multiplicity gravity amplitudes, and furthermore imply that only the helicity-conserving four-dimensional gravity amplitudes are nonvanishing upon dimensional reduction to $D=3$. This holds independent of the amount of supersymmetry. It is easy to confirm that the first non-trivial case, the four-point amplitude, is identical for the two- and three-algebra double-copy constructions. Given this, on-shell recursion~\cite{3dDualConformal}  can be used to show the equivalence of the two double- copy constructions~\cite{Future}. We demonstrate that this result is valid for $\mathcal{N}=6,4,2$ CSm theories, leading to $\mathcal{N}=12,10, 8$ supergravity theories.

We give a brief discussion of three-algebra based~\cite{McLoughlin} color-kinematics duality~\cite{BCJ}. A three-algebra is constructed via a triple product, antisymmetric in the first two entries, and four-indexed structure constants~\cite{N=8},
 \eq
 [T^a,T^b;\bar{T}^{\bar{c}}]=f^{ab\bar{c}}_{\phantom{abc}d}\,T^d\,.
 \eqe
The structure constants satisfy the fundamental identity
\eq
f^{ab\bar{c}}_{\phantom{abc}l}\,f^{dl\bar{e}\bar{g}}+
f^{ba\bar{e}}_{\phantom{abc}l}\,f^{dl\bar{c}\bar{g}}+
f^{*\bar{c}\bar{e}b}_{\phantom{abcd}\bar{l}}f^{da\bar{l}\bar{g}}+
f^{*\bar{e}\bar{c}a}_{\phantom{abcd}\bar{l}}f^{db\bar{l}\bar{g}}
=0\,,
\label{fund}
\eqe
where indices are raised/lowered by the metric $h^{a\bar{b}}={\rm Tr}(T^a\bar{T}^{\bar{b}})$. Imposing the structure constants to be real and totally antisymmetric leads to the $\mathcal{N}=8$ BLG theory. Relaxing the antisymmetry constraint leads to $\mathcal{N}=6,5,4$ theories~\cite{N=6,N=5}. 

Tree amplitudes of superconformal CSm theories are naturally represented by quartic diagrams, at $m$ points
\eq
\mathcal{A}_m=i\Big( \frac{2\pi}{k}\Big)^{\frac{m-2}{2}}\sum_{i\in {\rm quartic}}\frac{n_ic_i}{\prod_{\alpha_i}s_{\alpha_i}}  \,,
\label{quarticForm}
\eqe
where $i$ label the diagrams, $\alpha_i$ label propagators, and $k$ is the level. The color factors $c_i$ are constructed by dressing each quartic vertex by four-index structure constants. The ``kinematic numerators'', $n_i$, are (nonlocal) functions that encode the remaining state dependence. 

Following the color-kinematics duality for sYM theory, one can impose a similar duality between the kinematic numerators and color factors of the quartic diagrams. One requires that the numerators satisfy the same symmetries and identities as the color factors, schematically
\eqa
\label{dualityEq}
c_i\rightarrow -c_i~~&\Leftrightarrow&~~n_i\rightarrow -n_i\,\\
\nonumber c_i+c_j+c_k+c_l=0~~&\Leftrightarrow&~~n_i+n_j+n_k+n_l=0\,.
\eqae
The second line signifies the fundamental identity or generalized Jacobi identity. The double-copy principle state that once duality-satisfying numerators are found, the three-dimensional supergravity amplitude is given by 
\eq
\mathcal{M}_m=i\Big(\frac{\kappa}{2}\Big)^{m-2} \sum_{i\in {\rm quartic}}\frac{n_i\tilde{n}_i}{\prod_{\alpha_i}s_{\alpha_i}}\,,
\label{Sugra}
\eqe
where $\kappa$ is the gravity coupling. The $n_i,\tilde{n}_i$ may be identical or distinct CSm numerators depending on the theory under consideration. The formula is valid if at least one of the two sets of numerators satisfy the duality~\cite{BCJ2,Square}.

This brings us to our main equation: stating the equivalence of the three-dimensional supergravity amplitudes obtained from either two-algebra or three-algebra constructions. Suppressing $i(\kappa/2)^{m-2}$ factors, we have
\eq
\mathcal{M}_m~\,=  \hskip -0.2cm \mathop{\sum_{j\in {\rm cubic}}}_{N_j \in \text{2-algebra}} \hskip -0.2cm \frac{N_j\tilde{N}_j}{\prod_{\beta_j}s_{\beta_j}}~\,=  \hskip -0.2cm \mathop{\sum_{i\in {\rm quartic}}}_{n_i \in \text{3-algebra}} \hskip -0.1cm \frac{n_i\tilde{n}_i}{\prod_{\alpha_i}s_{\alpha_i}}  \,,
\label{MasterIdentity}
\eqe
where $N_j, \tilde{N}_j$ are sYM numerators of cubic graphs and $n_i, \tilde{n}_i$ are CSm numerators of quartic graphs. $N_j,n_i$ satisfy the kinematic two- and three-algebra, respectively. The particular theories encoded by these numerators need to be properly identified, examples are given in \tab{DCconstructions}.  Obtaining the sYM numerators is explained in ref.~\cite{BCJ}, so it will not be discussed here. The first equality in \eqn{MasterIdentity} has been proven~\cite{BCJ,Square}, and the second equality is the topic of the remainder of this paper.

Eq.~\ref{MasterIdentity} implies that ${\cal M}_4$ is a product of (color-stripped) CSm amplitudes: drooping all couplings, ${\cal M}_4=A_4 \tilde A_4$. At six points, the appearance of propagators and the fundamental identity results in an intriguing interplay. Considering the ${\cal N}\le6$ CSm theories one can form nine distinct color factors $c_i$ given that odd legs $1,3,5$ have bared color indices $\bar a$. These nine contributing diagrams can be identified from their three-particle channels
$$s_i\equiv(
s_{123},\;s_{126},\;s_{134},\;s_{125},\;s_{146},\;s_{136},\;s_{145},\;s_{124},\;s_{156})\,.$$
The duality implies that the number of independent color factors and independent numerator factors is one to one;  $p=5$ at six points. Expressing the five color-stripped amplitudes in terms of the independent numerators gives
\eq
A_{(i)}=\sum_{j=1}^{p}\Theta_{ij}n_j  \,.
\label{ThetaDef}
\eqe
This defines the $\Theta$ matrix, which is comprised of sums of products of propagators with $\pm1$ coefficients. At four points this matrix is trivial, $\Theta=1$, and at six points $\Theta$ is a five-by-five matrix, given by
\eq
\left(\begin{array}{ccccc}
\frac{1}{s_{1}} & \frac{1}{s_{2}}+\frac{1}{s_{9}} & \frac{1}{s_{9}} & -\frac{1}{s_{9}} & 0 \\ 
\frac{1}{s_{8}} & -\frac{1}{s_{8}} & \frac{1}{s_{3}} & \frac{1}{s_{4}} +\frac{1}{s_{8}}& 0 \\ 
\frac{1}{s_{7}} & -\frac{1}{s_{7}} & -\frac{1}{s_{6}}-\frac{1}{s_{7}} & \frac{1}{s_{6}}+\frac{1}{s_{7}} &\frac{1}{s_{5}}+ \frac{1}{s_{6}}+\frac{1}{s_{7}} \\
0 & -\frac{1}{s_{9}} & -\frac{1}{s_{3}}-\frac{1}{s_{9}} & \frac{1}{s_{9}}& -\frac{1}{s_{5}} \\
0 & -\frac{1}{s_{2}} & \frac{1}{s_{6}} & -\frac{1}{s_{4}}-\frac{1}{s_{6}} & -\frac{1}{s_{6}}
\end{array}\right) \,,
\eqe 
where the five independent color-ordered amplitudes $A_{(i)}$ are chosen to be $ A(\bar1,2,\bar3,4,\bar5,6)$, $A(\bar1,4,\bar3,6,\bar5,2)$, $A(\bar1,6,\bar3,2,\bar5,4)$, $A(\bar1,4,\bar3,2,\bar5,6)$ and $A(\bar1,6,\bar3,4,\bar5,2)$. The five independent numerators are $n_1,n_2,n_3,n_4,n_5$.

Naively, inverting $\Theta$ would give duality-satisfying numerators. However, the matrix $\Theta$ is not invertible; at six points it has only rank 4. Although counterintuitive, this is a desirable property. It implies that at least one $n_i$ corresponds to ``pure gauge"  and can thus be set to any convenient value while still obtaining the correct amplitude.   Using this property one can work out non-trivial relations between color-ordered amplitudes.
As a side remark, we note that the matrix $\Theta$ has full rank if one employs either $D>3$ or off-shell momenta. The former is in contrast to the sYM case, where the corresponding $\Theta$ cannot be inverted in any space-time dimension.

For the six-point case, one obtains a single amplitude relation via the kernel
\eq
{\rm Ker}(\Theta^{T})\cdot A =\sum_{i=1}^5C_{ik}A_{(i)}=0 \,,
\eqe  
where $C_{ik}= (-1)^{i+k} \, M_{ik}$ is the $(i,k)$ cofactor, and $M_{ik}={\rm Det}(\Theta_{\hat i \hat k})$ is the $(i,k)$ minor of the matrix $\Theta$. Note that this formula is equivalent to replacing the $k$-th column of $\Theta$ by $A_{(i)}$ and then demanding that the resulting matrix has zero determinant. All choices for $k$ give the same relation.
Up to an overall irrelevant factor, the coefficients $C_{ik}$ are degree-four polynomials in $s_i$. 

Setting, say, $n_5$ to be zero, and omitting, say, $A_{(5)}$, we can now invert the reduced matrix $\Theta_{ij}$ to express $n_1,n_2,n_3,n_4$ in terms of color ordered $\mathcal{N}=6,4,2$ CSm  amplitudes. Inserting the resulting $n_i$ into \eqn{MasterIdentity}, with appropriate pairing, we have explicitly checked that one obtains correct $\mathcal{N}=12,10,8$ supergravity amplitudes. 

%%%%%%%%%%%%%%%%%%%%%%%%%%%%%%%%%%%%%%%
 \subsection{Explicit amplitudes}
 %%%%%%%%%%%%%%%%%%%%%%%%%%%%%%%%%%%%%%
%
%%%%%%%% TABLE %%%%%%%%%%%
\def\hs{\hskip .2 cm \null }
\begin{table*}
\caption{Examples of explicitly-confirmed double-copy constructions of supergravity theories with half-maximal or more supersymmetry. $\mathcal{N}=8$ CSm theory has $16$ states while $\mathcal{N}=6,4,2$ CSm have $(8,\bar{8})$, $(4,\bar{4})$, $(2,\bar{2})$ states in the (chiral, antichiral) multiplet, respectively. For $\mathcal{N}=8,4,2,0$ sYM we use the (four-dimensional) state counts $16,8,4,2$. The single state ``1'' denotes pure $D=3$ YM, or single-scalar CSm theory, in the respective columns. Here $n$ counts matter multiplets.}
\label{DCconstructions} 	
\vskip .1 cm
\begin{tabular}{||c|r|l|c||}
\hline 
SG theory & \, CSm$_{\rm L}$$\times$CSm$_{\rm R}=$ supergravity ~~~~~~~~ & \, sYM$_{\rm L}$$\times$sYM$_{\rm R}=$ supergravity \, & coset \\
\hline
 ${\cal N}=16$ & $16^2=256 \hskip 2.275cm$  & $\hskip 1.63cm 16^2=256$ & E$_{8(8)}$/SO(16)\\
\hline
 ${\cal N}=12$ &  $8^2+\bar{8}^2\,=\,16\times (4+\bar{4})\;=128$ & $ \hskip 1.22cm 16\times 8=128$   &~ E$_{7(-5)}$/SO(12)$\otimes$SO(3)\\
\hline
${\cal N}=10$ &   $8 \times 4+\bar{8}\times \bar{4}\,=\,16\times (2+\bar{2})\;=64\phantom{1}$  & $\hskip 1.22cm 16\times 4=64$ & E$_{6(-14)}$/SO(10)$\otimes$SO(2)\\
\hline
 ${\cal N}=8$, $n=2$& $\,4^2+\bar{4}^2=8 \times 2+\bar{8}\times \bar{2}=32\phantom{1}$ 
  & $\hskip 1.22cm 16\times 2=32$  & SO(8,2)/SO(8)$\otimes$SO(2)\\
\hline
${\cal N}=8$, $n=1$&  $16 \times 1=16\phantom{1}\hskip 2.275cm$
  & $\hskip 1.22cm 16\times 1=16$  & SO(8,1)/SO(8)\\
  \hline
\end{tabular}
\vskip -0.1 cm
\end{table*}
%%%%%%%%%%%%%%%%%%%%%%%
%
In order to clarify the bookkeeping details for the various supergravities we now list the four-point double-copy amplitudes. The maximal $\mathcal{N}=16$ supergravity case is a square of the $f^{abcd}$-stripped BLG  $\mathcal{N}=8$ amplitude~\cite{McLoughlin}. Suppressing couplings and $i$'s henceforth, we have
\eqa
\nonumber {\cal M}_4^{{\cal N}=16}&=&\frac{\delta^{(16)}(\sum_{i}\lambda^\alpha\eta^I_i)}{s_{12}s_{23}s_{31}}\\
&=&\left(\frac{\delta^{(8)}(\sum_{i}\lambda^\alpha\eta^I_i)}{\spa{1}.{2}\spa{2}.{3}\spa{3}.{1}}\right)^2=(A_4^{{\cal N}=8})^2\,,
\label{ANeq16}
\eqae
where the square of the fermionic delta function is understood as a tensor product for the Grassmann-valued $\eta^{I}_i$, with R-charge indices $I\in(1,\ldots, {\cal N}/2)$ for each theory.  We use standard spinor products $\spa{i}.{j}=\lambda^\alpha\lambda^\beta \eps_{\alpha \beta}=\sqrt{s_{ij}}$. With maximal supersymmetry the on-shell states form a single multiplet, but for fewer supersymmetries the states split into two multiplets $\Phi^{{\cal N}}$ and $\overline \Phi^{{\cal N}}$ similar to that defined in refs.~\cite{Till,YT1}. These are chiral and antichiral multiplets of ${\rm SU}({\cal N}/2) \subset {\rm SO}({\cal N})$. Here we take ${\cal N}$ to be even; theories with odd ${\cal N}$ we leave for future work. Taking legs $1$ and $3$ to be antichiral multiplets, the $\mathcal{N}=12$ amplitude is  
\eq
{\cal M}_4^{{\cal N}=12}(\bar{1},2,\bar{3},4)=(A_4^{{\cal N}=6})^2=\left(\frac{\delta^{(6)}(\sum_{i}\lambda^\alpha\eta^I_i)}{\spa{1}.{2}\spa{2}.{3}}\right)^2\,,
\label{ANeq12}
\eqe
where $A_4^{{\cal N}=6}=A_4^{{\cal N}=6}(\bar{1},2,\bar{3},4)$ is the color-stripped four-point amplitude~\cite{Till} of the $\mathcal{N}=6$ theory constructed by Aharony, Bergman, Jafferis and Maldacena (ABJM) \cite{ABJM}. One may also construct the $\mathcal{N}=12$ amplitude as a heterotic double copy $A_4^{{\cal N}=8} \times A_4^{{\cal N}=4}$. This gives the correct result, as explicitly verified up to six points, even though the structure constants and hence the numerators of the two theories obey different symmetries. For $\mathcal{N}=10$ supergravity the four-point amplitude is given by
 \eqa
 {\cal M}_4^{{\cal N}=10}(\bar{1},2,\bar{3},4)= \frac{\delta^{(10)}(\sum_{i}\lambda^\alpha\eta^I_i)\spa{1}.{3}}{\spa{1}.{2}^2\spa{2}.{3}^2}\,.
 \label{ANeq10}
 \eqae
It can be constructed as heterotic double copies $A_4^{{\cal N}=8} \times A_4^{{\cal N}=2}(\bar{1},2,\bar{3},4)$ or $A_4^{{\cal N}=6}(\bar{1},2,\bar{3},4) \times A_4^{{\cal N}=4}(\bar{1},2,\bar{3},4)$.

As discussed in ref.~\cite{deWit:1992up} all supergravity theories with $\mathcal{N}>8$ supersymmetry are believed to be unique, while beginning with $\mathcal{N}=8$ one can have $n$ matter multiplets, corresponding to $16 n$ states. The dimensional reduction of  pure half-maximal $D=4$ supergravity corresponds to $\mathcal{N}=8$ with $n=2$; the four-point amplitude is
\eq
{\cal M}_{4,n=2}^{{\cal N}=8}(\bar{1},2,\bar{3},4)=(A_4^{{\cal N}=4})^2\hskip-0.5mm=\left(\frac{\delta^{(4)}(\sum_{i}\lambda^\alpha\eta^I_i)\spa{1}.{3}}{\spa{1}.{2}\spa{2}.{3}}\right)^2,
\label{ANeq8}
\eqe
with $A_4^{{\cal N}=4}=A_4^{{\cal N}=4}(\bar{1},2,\bar{3},4)$. One can also write this as $A_4^{{\cal N}=6} \times A_4^{{\cal N}=2}$. For $n=1$, the four-point amplitude is 
\eq
{\cal M}_{4,n=1}^{{\cal N}=8}=\frac{1}{2}\frac{\delta^{(8)}(\sum_{i}\lambda^\alpha\eta^I_i)(s_{12}^2+s_{23}^2+s_{13}^2)}{\spa{1}.{2}^2\spa{2}.{3}^2\spa{1}.{3}^2}\,.
\label{ANeq8pure}
\eqe
This is given by a direct product of $A_4^{{\cal N}=8}\times A_4^{{\cal N}=0}$, where ${\cal N}=0$ denotes CS theory plus a single minimally-coupled scalar. We summarize these results in \tab{DCconstructions}.

%%%%%%%%%%%%%%%%%%%%%%%%%%%%%%%%%%%%%%%
 \subsection{Vanishing of odd-multiplicity amplitudes}
 %%%%%%%%%%%%%%%%%%%%%%%%%%%%%%%%%%%%%%
%
%
The validity of double-copy formula  (\ref{MasterIdentity}) for all multiplicities can be proven~\cite{Future} utilizing on-shell recursion formulas for the amplitudes. The proof is similar to that of the two-algebra double-copy relations given in ref.~\cite{Square}. It relies on having a well-behaved three-dimensional on-shell recursion formula. One such was given for the case of  $\mathcal{N}=6$ CSm theory in ref.~\cite{3dDualConformal}. This recursion extends straightforwardly for theories with only even-multiplicity S-matrix elements, and for amplitudes that vanish at large complex deformation. The latter property can be shown to be inherited from four-dimensional supergravity amplitudes; here, we show the former property for supergravity using R-symmetry arguments.

First we consider the constraint from R-symmetry in four dimensions (see~\cite{NBIRelationsRsym} for similar discussion). Through the KLT relations, that is, the two-algebra double-copy formula, maximally supersymmetric $\mathcal{N}=8$ gravity  inherits an enhanced SU(8) R-symmetry. This includes the following U(1) generator
\eqa
\;\;R=\sum_{i=1}^m\eta_i^{I_{\rm L}}\frac{\p}{\p \eta_i^{I_{\rm L}}}-\teta_i^{I_{\rm R}}\frac{\p}{\p \teta_i^{I_{\rm R}}}\,,
\label{SU8R}
\eqae
where $I_{\rm L},I_{\rm R}\in 1,\cdots, 4$. Applied to the amplitude, the generator $R$ counts the $\eta$ degree minus the $\teta$ degree, or, as the $\eta$'s are charged under helicity, the difference of helicity weight between left and right amplitudes. Denoting the KLT map as ${\cal M}=K[\mathcal{A}_{\rm L},\mathcal{A}_{\rm R}]$, R-symmetry invariance thus requires that the two $\mathcal{N}=4$ sYM amplitudes must have the same helicity weight:
\eq
K[\mathcal{A}^{{\rm N}^{k}{\rm MHV}}_{\rm L},\mathcal{A}^{{\rm N}^{ k'}{\rm MHV}}_{\rm R}]~~\left\{\begin{array}{c}=0\;\;{\rm for}\;k\neq k' \\ \neq 0\;\;{\rm for}\;k=k'\end{array}\right. \,,
\label{kconstraint}
\eqe
where N$^{k}$MHV stands for (next-to-)$^{k}$maximally-helicity-violating amplitude. Note that since one can consistently truncate supersymmetry on both sides of the KLT formula to obtain reduced supersymmetric theories, the above condition is valid for all tree-level pure (super) gravity amplitudes. 

Reducing four-dimensional $\mathcal{N}=q$ supergravity to three dimensions, one obtains an enhanced SO($2q$) symmetry. The SO($2q$) generators are built out of quadratic forms $\sim \eta^2$, $\eta \p_\eta$ and $(\p_\eta)^2$, among these one can identify the U(1) generator $Y=Y_{\rm L}+Y_{\rm R}$, where
\eq
Y_{\rm L}=\frac{1}{2}\left(\sum_{i=1}^m\eta^{I_{\rm L}}_i\frac{\p}{\p \eta^{I_{\rm L}}_i}\right)-m\,,
\eqe
and $Y_{\rm R}$ is similarly defined in terms of the $\teta$ variables. As $R=2(Y_{\rm L}-Y_{\rm R})$ it follows that $Y_{\rm L}$ and $Y_{\rm R}$ must vanish individually. This freezes the number of $\eta$'s, or $\teta$'s, to be $2m$, corresponding to helicity weight $m$. Additionally, any $D=4$ sYM amplitude carries overall helicity weight $-m$ not accounted for by the $\eta$'s (cf. Park-Taylor denominator). Thus, in total, only helicity-conserving Yang-Mills amplitudes -- present exclusively at even multiplicity -- can give nonvanishing gravity amplitudes in the KLT or double copy formula. Equivalently, four-dimensional gravity amplitudes have nonvanishing three-dimensional descendant only for helicity-conserving configurations. We have checked this explicitly for all N$^k$MHV sectors of graviton tree amplitudes up to 10 points. Unitarity of the S-matrix suggests that the vanishing of odd-multiplicity and helicity non-conserving amplitudes continues at loop level, however, the need for regularization of potential UV and IR divergences may complicate the details.

In conclusion, we have shown that the three-algebra based double-copy formula relates a large class of CSm amplitudes to $D=3$ supergravity amplitudes. Remarkably, the same gravity amplitudes can be obtained from two-algebra based double-copy of sYM amplitudes, as been previously shown~\cite{BCJ, Square}. This is striking as CSm and sYM amplitudes have conspicuously distinct properties, such as (non)vanishing odd-multiplicity S-matrix elements. We have also clarified that amplitude relations arise from the fact that the matrix $\Theta_{ij}$ is of lower rank, which is only true for $D=3$. It would be interesting if the resulting amplitude relations have a string-theory explanation, as was the case for sYM~\cite{monodromy}. Finally, we note that the existence of a three-algebra double-copy formula may have intriguing consequences for the UV behavior of three-dimensional supergravity, which, just as in four dimensions, is nonrenormalizable by naive power counting.  Loop-level numerators that satisfy three-algebra color-kinematics must necessarily be nonlocal, due to the existence of soft poles in the four-point amplitudes~(\ref{ANeq16})-(\ref{ANeq8pure}). Such a nonlocal behavior, at each four-point vertex, has the potential to improve the naive UV power counting of supergravity. Viewing three-dimensional supergravity as a decoupling limit of string theory~\cite{Green}, also suggests a better UV behavior.  Together, these clues suggest that a construction of explicit duality-satisfying loop-level numerators could advance our understanding of the detailed UV structure of gravity theories.

\vskip .3 cm 

We thank BIRS and the Banff Centre for their hospitality during the workshop ``The Geometry of Scattering Amplitudes" which partly inspired this work. We also thank T. Bargheer, Z. Bern, J. J. Carrasco, M. Green, N. Lambert, S. Lee, T. McLoughlin and D. O'Connell for fruitful discussions. This research was supported by the US DoE grant DE-SC0007859, and by the European
Research Council under Advanced Investigator Grant ERC-AdG-228301.

\end{document}